%% file: Main_Manuscript_arXiv.tex
\documentclass[12pt]{article}
\usepackage[T1]{fontenc}
\usepackage[utf8]{inputenc}
\usepackage{lmodern}
\usepackage{microtype}
\usepackage[english]{babel}

\usepackage[a4paper,margin=2.5cm]{geometry}
\usepackage{setspace}
\usepackage{amsmath,amssymb,amsfonts,amsthm,mathtools}
\usepackage{bm}

\usepackage{booktabs}
\usepackage{tabularx}
\usepackage{array}
\usepackage{multirow}
\usepackage{enumitem}

\usepackage{graphicx}
\usepackage{caption}
\usepackage{subcaption}
\usepackage{float}
\usepackage[section]{placeins}

\usepackage{mdframed}
\newmdenv[
  skipabove=8pt,
  skipbelow=8pt,
  linewidth=0.6pt,
  roundcorner=4pt,
  innerleftmargin=8pt,
  innerrightmargin=8pt,
  innertopmargin=6pt,
  innerbottommargin=6pt
]{naivereadingbox}

\usepackage[dvipsnames]{xcolor}

\usepackage[authoryear,round]{natbib}

\usepackage{tikz}
\usetikzlibrary{arrows.meta,calc,positioning}
\usepackage{pgfplots}
\usepgfplotslibrary{groupplots}
\pgfplotsset{compat=1.18}

\usepackage{indentfirst}

\newcommand{\BiplotKeepEvery}{1}
\newcommand{\EnvelopeKeepEvery}{2}

\newcounter{biplotpt}
\newcommand{\MaybePlotBiplotPoint}[1]{\stepcounter{biplotpt}\pgfmathtruncatemacro{\biplotptmod}{mod(\value{biplotpt},\BiplotKeepEvery)}\ifnum\biplotptmod=0\relax
    #1\fi
}

\newcounter{bootseg}
\newcommand{\MaybePlotEnvelope}[1]{\stepcounter{bootseg}\pgfmathtruncatemacro{\bootsegmod}{mod(\value{bootseg},\EnvelopeKeepEvery)}\ifnum\bootsegmod=0\relax
    #1\fi
}

\usepackage{fancyhdr}
\newtheorem{theorem}{Theorem}

\newtheorem{proposition}{Proposition}

\theoremstyle{definition}
\newtheorem{definition}{Definition}

\theoremstyle{remark}
\newtheorem{remark}{Remark}

\DeclareMathOperator{\Corr}{Corr}

\usepackage{hyperref}
\usepackage[nameinlink,noabbrev]{cleveref}
\newcommand{\doi}[1]{\href{https://doi.org/#1}{doi:#1}}
\hypersetup{
  colorlinks=true,
  linkcolor=blue,
  citecolor=blue,
  urlcolor=blue
}
\crefname{definition}{Definition}{Definitions}
\Crefname{definition}{Definition}{Definitions}
\crefname{theorem}{Theorem}{Theorems}
\Crefname{theorem}{Theorem}{Theorems}
\crefname{corollary}{Corollary}{Corollaries}
\Crefname{corollary}{Corollary}{Corollaries}
\newcommand{\SMone}{Supplementary Material SM1}
\newcommand{\SMtwo}{Supplementary Material SM2}
\newcommand{\SMthree}{Supplementary Material SM3}
\newcommand{\SMfour}{Supplementary Material SM4}
\newcommand{\SMfive}{Supplementary Material SM5}

\newcommand{\SMfourTitle}{\emph{Spectral Identifiability, Projection Diagnostics, and Additional Controls}}

\newcommand{\MainRepo}{main-manuscript reproducibility materials}

\title{Claim-Specific Admissibility of PCA Biplot Interpretations: Target Alignment, Spectral Identifiability, and Projection Adequacy}
\author{\parbox{0.92\textwidth}{\centering Luiz Roberto Martins Pinto$^{1,*}$ and Carlos Tadeu dos Santos Dias$^{2}$\\[0.7em]
\small $^{1}$Retired professor, Department of Exact and Technological Sciences, State University of Santa Cruz (UESC),\\
\small Campus Soane Nazar\'e de Andrade, Rodovia Jorge Amado, Km 16, Salobrinho,\\
\small Ilh\'eus, Bahia 45662-900, Brazil\\[0.35em]
\small $^{2}$Retired professor, Department of Exact Sciences, Luiz de Queiroz College of Agriculture (ESALQ),\\
\small University of S\~ao Paulo (USP), Avenida P\'adua Dias 11, Piracicaba,\\
\small S\~ao Paulo 13418-900, Brazil\\[0.5em]
\small $^{*}$Corresponding author: Luiz Roberto Martins Pinto; \href{mailto:luizrobertomp@gmail.com}{luizrobertomp@gmail.com}\\
\small ORCID: Luiz Roberto Martins Pinto, \href{https://orcid.org/0000-0003-3156-8742}{0000-0003-3156-8742};\\
\small Carlos Tadeu dos Santos Dias, \href{https://orcid.org/0000-0003-1015-1761}{0000-0003-1015-1761}}
}
\date{}
 
\begin{document}
\maketitle
\thispagestyle{plain}

\begin{abstract}
Unit-variance standardisation is often applied routinely before principal component analysis (PCA), although it replaces covariance geometry by correlation geometry. A resulting biplot may be computed correctly yet fail to support a scientific interpretation about the original-scale phenomenon. We therefore make the scientific statement, rather than the decomposition alone, the unit of methodological assessment. A PCA-biplot claim is representationally well posed only when the declared scientific target justifies the operator analysed, the invoked axis or invariant subspace is identifiable at the stated level, and the displayed projection preserves the prespecified relationships within a substantively justified tolerance; failure of any condition makes the claim ill posed for the stated interpretation. A projector formulation yields basis-invariant diagnoses for repeated-eigenvalue blocks, and a residual-Gram bound quantifies pairwise error from omitted coordinates. Six controlled population scenarios provide exact representational truth, including zero target association with arbitrary projected angles, collapse of a full-space $60^\circ$ relationship to $0^\circ$, and non-identifiable named axes. The contribution is not another reminder that scaling matters: it establishes that computational correctness is necessary but not sufficient for scientific interpretability and provides a formal procedure for retaining, reformulating, qualifying, or rejecting a PCA-biplot statement. A real-data illustration and bootstrap extension are provided as supplementary material.
\end{abstract}

\vspace{0.5em}
 
\section{Introduction}
\label{sec:introduction}

Principal component analysis (PCA) is one of the most widely used tools in multivariate
analysis, serving both as a linear method for dimension reduction and as a basis for graphical
exploration. Since the foundational work of \citet{Hotelling1933}, PCA has become routine
across biology, agriculture, medicine, ecology, chemometrics, and the social sciences; see,
among many others, \citet{Jolliffe2002}, \citet{AbdiWilliams2010}, and
\citet{JolliffeCadima2016}. In applied work, PCA is rarely used only to compress dimension. It
is also expected to reveal dominant patterns, organise observations in a reduced space, suggest
latent gradients, and support geometric readings of inter-variable association through loadings
and biplots.

This graphical role, long recognised in exploratory data analysis \citep{Tukey1977}, gives the present problem practical importance. PCA biplots are often used
not merely as descriptive summaries, but to motivate scientific narratives about groups of
variables, trade-offs, co-movement, latent gradients, or biological and technological mechanisms.
If such narratives are attached to a display whose target geometry differs from the phenomenon
being discussed, the problem is not a minor graphical imperfection. It is a risk of assigning
substantive meaning to angles, vector lengths, projections, and apparent separations that are not
supported by the operator actually represented.

A central modelling choice in PCA is the scale on which variation is defined. Major multivariate textbooks and methodological articles have long recommended standardisation when variables have heterogeneous units or marginal dispersions, when arbitrary changes of measurement unit should not alter the analysis, or when equal marginal influence is substantively intended \citep{Wold1987,Jackson1991,Jolliffe2002,Rencher2002,BroSmilde2003,Manly2005,VandenBerg2006,JohnsonWichern2007,AbdiWilliams2010,Hair2010,JolliffeCadima2016,Forkman2019,Brereton2025}. This legitimate, originally conditional guidance helps explain why unit-variance scaling has become conventional, and in some fields nearly default, applied practice. It also establishes that standardisation changes the analysed operator: PCA of z-scores diagonalises the correlation matrix rather than the raw covariance matrix. Consequently, the resulting biplot represents standardised linear co-movement, not the original covariance geometry, and must be interpreted on that scale. \SMone{} documents two complementary evidence layers---the canonical teaching of standardisation and a structured illustrative sample of recent applied PCA-biplot practice---providing the historical and applied evidence that motivates the problem, supports its relevance, and helps delimit the manuscript's novelty.

The unresolved issue is therefore not that the distinction between covariance-based and correlation-based PCA has gone unrecognised, nor that biplot geometry lacks established interpretive rules, including recent work devoted specifically to correlation-matrix biplots \citep{Graffelman2025}. Rather, these traditions are not usually integrated through a claim-specific assessment of whether a scientific interpretation is aligned with the matrix actually diagonalised, refers to an identifiable spectral object, and is adequately preserved in the displayed projection. The present manuscript addresses that combined representational question.

The practical question motivating this manuscript is consequently simple but consequential:
what happens when z-score or unit-variance standardisation is applied routinely before PCA and
the resulting biplot is interpreted as though it still represented the original raw-scale
phenomenology? The relevant issue is not whether correlation-PCA is mathematically valid. It is
whether the correlation-scale geometry produced by standardisation is the geometry required by
the scientific question. Standardisation is defensible when the question is explicitly formulated
on the correlation scale; it may render the intended interpretation ill posed when the question
still concerns raw-scale covariance, absolute variation, or another feature of the original
measurement scale. Throughout this paper, we work with theoretical population operators and
restrict attention to linear association, the notion directly encoded by covariance, Pearson
correlation, and classical PCA geometry.

To analyse this problem precisely, we separate four questions that are often conflated. The first
is \emph{target change}: unit-variance standardisation replaces the covariance target $\Sigma$
with the correlation target $R$. The second is \emph{target alignment}: the new target may be
legitimate for one scientific question yet inappropriate for another. The third is \emph{spectral
identifiability}: repeated or nearly repeated eigenvalues may leave individual principal axes
non-identifiable or unstable even when the target matrix itself is fully defined. The fourth is
\emph{projection adequacy}: a low-dimensional biplot may reproduce only part of the relevant
full-dimensional inner-product structure. These questions must be diagnosed separately. A
change from $\Sigma$ to $R$ is not, by itself, evidence of spectral instability, and a spectrally
stable correlation-PCA biplot may still be scientifically misaligned with a raw-scale question.
Conversely, a scientifically appropriate target may contain repeated eigenvalues and therefore
fail to identify individual PC axes.

The paper develops a claim-specific admissibility framework rather than a new eigendecomposition. Admissibility is the formal criterion; representational well-posedness is the resulting claim-specific status. Its methodological contribution has three integrated parts. First, it makes the scientific target an explicit component of the claim and therefore treats target alignment as a documented modelling requirement rather than an unspoken preprocessing choice. Second, it formulates spectral interpretation through projectors, so that statements about repeated-eigenvalue subspaces are invariant to arbitrary rotations of their internal bases. Third, it measures projection adequacy for the relationships actually invoked by the claim and supplies a residual-Gram bound for omitted-coordinate error. Six controlled population scenarios make these conditions exactly verifiable and demonstrate that apparently persuasive biplot geometry can contradict, distort, or fail to identify the phenomenon invoked by the interpretation.

\subsection*{Methodological contribution relative to existing PCA and biplot theory}

The contribution is not the isolated observation that scaling matters, that repeated eigenvalues destabilise axes, or that two-dimensional displays are approximate. Those facts are established. The methodological advance is to make a scientific PCA-biplot statement itself the object of assessment and to require three formerly separate considerations---target choice, spectral identifiability, and low-rank representation---to license that statement jointly. Computational correctness, high retained inertia, and visual coherence are therefore necessary aids to interpretation, but none establishes that the displayed geometry represents the phenomenon named in the conclusion.

The proposed specification records the phenomenon and scale of interest, the operator that carries that phenomenon, the spectral object to which the language refers, the displayed component set, the relationships on which the statement depends, and the allowable representation error. It returns a practical decision: retain the claim, reformulate it at subspace level, qualify it as plane-specific or approximate, redirect it to another component set, or reject it because the represented operator answers a different scientific question. The issue is whether the particular sentence written from a particular display is licensed by its target, identifiable spectral object, and projection.

The paper is organised as follows. \Cref{sec:framework} develops the population admissibility framework. \Cref{sec:controlled-contrasts} gives controlled population contrasts before and after unit-variance standardisation. \Cref{sec:discussion} discusses methodological implications and limitations, and \cref{sec:conclusion} concludes. The supplementary materials document the historical motivation (\SMone{}), the biplot atlas and diagnostics (\SMtwo{}), a positive correlation-PCA control (\SMthree{}), expanded spectral and projection diagnostics (\SMfour{}), and the empirical Iris application with bootstrap assessment (\SMfive{}). Results, figures, scripts, and machine-readable outputs directly supporting the main text are supplied in the \MainRepo{}; supplement-specific scripts and outputs accompany the corresponding supplementary materials.

  \section{Target Geometry and Representational Well-Posedness}
\label{sec:framework}

\subsection{The target shift induced by unit-variance standardisation}
\label{subsec:target-shift}

Let $X\in\mathbb R^p$ be a random vector with positive marginal variances, population covariance matrix $\Sigma$, and $D=\operatorname{diag}(\sigma_1,\ldots,\sigma_p)$. Unit-variance standardisation replaces $X$ by $Z=D^{-1}(X-\mathbb EX)$ and therefore replaces the population operator $\Sigma$ by
\[
R=D^{-1}\Sigma D^{-1}.
\]
For a sample $X_1,\ldots,X_n$, the corresponding plug-in operators are $\widehat\Sigma$ and $\widehat R=\widehat D^{-1}\widehat\Sigma\widehat D^{-1}$. Covariance-PCA and correlation-PCA are therefore estimators of different population eigensystems.

\begin{proposition}[Population and sample target shift]
PCA of the population-standardised vector $Z$ diagonalises $R$, whereas PCA of sample z-scores diagonalises $\widehat R$. If $\Sigma$ is diagonal with unequal positive diagonal entries, its ordered eigenvectors are the coordinate axes up to sign, while $R=I_p$ identifies neither an individual principal axis nor a proper leading principal subspace. The analogous statement holds for a diagonal sample covariance matrix with unequal positive entries.
\end{proposition}

\noindent The result follows from $\operatorname{Cov}\{D^{-1}(X-\mathbb EX)\}=D^{-1}\Sigma D^{-1}$, with the same calculation for sample plug-in operators; the spectral statements are immediate for diagonal matrices and for $I_p$. The consequence is not only algebraic. Multiplying by $D^{-1}$ removes marginal variances from the represented operator and replaces the raw inner-product geometry of variables by a unit-variance geometry. Thus the biplot obtained after z-scoring may be perfectly computed while representing a different scientific object from the one invoked by a statement about raw-scale variation or covariance.

\begin{remark}[Exceptional agreement of spectral objects]
Covariance-PCA and correlation-PCA can share particular invariant subspaces under special structure, but such agreement must be established for the operators at hand. It is not implied by standardisation.
\end{remark}

The target change is mathematically legitimate. The methodological question is whether the correlation operator is required by the scientific question. A stable correlation-PCA can be scientifically misaligned with a raw-scale question, whereas a scientifically appropriate covariance-PCA can still contain non-identifiable axes.
 \subsection{Representational well-posedness}
\label{subsec:wellposedness}

A PCA biplot is a low-rank representation of the eigensystem of a specified symmetric positive semidefinite target matrix $M$ \citep{Gabriel1971,GowerHand1996,Greenacre2010}. Its angles, lengths, and projections inherit their meaning from that target, the biplot scaling, and the retained dimensions.

\begin{definition}[Scientific target declaration]\label{def:target}
A target declaration $T$ states the population quantity whose geometry is scientifically relevant and the invariance requirements that motivate it. Examples include raw-scale covariance $\Sigma$, correlation $R$, or another explicitly defined positive-semidefinite second-order operator. Target alignment is not inferred from the data alone: it is satisfied only when the declaration and its invariances justify the operator analysed.
\end{definition}

\begin{definition}[PCA-biplot claim]\label{def:claim}
A PCA-biplot claim is specified by
\[
\mathcal C=(T,M,\mathcal O,J,\mathcal B,\mathcal P,r,d,\tau).
\]
The first entries specify the scientific and geometric object under discussion: $T$ is the target declaration, $M$ is the population operator represented, $\mathcal O$ is the invoked spectral object, expressed as a simple-axis projector or a block spectral projector, $J$ is the displayed component block, and $\mathcal B$ is the biplot coordinate convention.

The remaining entries specify how the claim will be audited: $\mathcal P$ is the prespecified set of variable pairs or contrasts required by the claim, $r_{jk}(M)$ and $r_{jk}^{(J,\mathcal B)}(M)$ are the full and displayed relationships, $d$ is a non-negative discrepancy, and $\tau\geq0$ is a scientifically justified tolerance. The tuple is a bookkeeping device for the statement being audited, not an additional PCA model.
\end{definition}

\begin{definition}[Conditional population admissibility]\label{def:admissibility}
Conditional on the declared $(T,\mathcal B,\mathcal P,r,d,\tau)$, the claim $\mathcal C$ is population-admissible when:
\begin{enumerate}[label=(\roman*)]
\item \emph{target alignment}: the invariances and scale encoded by $T$ justify $M$;
\item \emph{spectral identifiability}: the projector $\mathcal O$ is uniquely determined by $M$ at the level invoked by the claim;
\item \emph{projection adequacy}:
\[
\Delta_J(\mathcal C)=\max_{(j,k)\in\mathcal P}
 d\!\left\{r_{jk}^{(J,\mathcal B)}(M),r_{jk}(M)\right\}\leq\tau.
\]
The inequality is the operational form of projection adequacy: it asks whether the relation actually read from the displayed plane is close enough to the corresponding relation in the declared target matrix for the specific claim under review.

\end{enumerate}
Failure of any condition identifies a distinct reason why the stated interpretation is not licensed. The word \emph{conditional} is essential: the framework audits a declared scientific specification; it does not infer the target or tolerance from the biplot.
\end{definition}

\paragraph{Claim-specific representational status.}
Conditional on the declared specification, a population-admissible claim is \emph{representationally well posed}; if any condition fails, it is \emph{representationally ill posed} for that interpretation. The status belongs to the claim, not to PCA, correlation-PCA, or the biplot as a whole: the same display may support one well-posed statement and fail to support another.

\begin{theorem}[Basis-invariant spectral representation and residual bound]\label{thm:projector}
Let $M=Q\Lambda Q^\top$ be positive semidefinite and let $J$ index a union of complete eigenspaces. Write $P_J=Q_JQ_J^\top$ and
\[
M_J=P_JMP_J=Q_J\Lambda_JQ_J^\top,\qquad H_J=M-M_J.
\]
Then: (a) $P_J$, $M_J$, and $H_J$ are invariant to every orthogonal change of basis within repeated-eigenvalue blocks; (b) $H_J$ is positive semidefinite; and (c), for every variable pair,
\[
\left|M_{jk}-(M_J)_{jk}\right|=|(H_J)_{jk}|
\leq\sqrt{(H_J)_{jj}(H_J)_{kk}}.
\]
Consequently, projector-based projection diagnostics are well defined even when individual axes inside a retained repeated block are not. This result is used to separate a legitimate subspace-level interpretation from an illegitimate named-axis interpretation within a repeated eigenspace.
\end{theorem}

The proof and the corresponding uniform pairwise screening corollary are given in \SMfour{} (\SMfourTitle{}). The Gram-matrix formulation is retained in the main text because it identifies the quantity that a biplot can and cannot display. The target matrix $M$ stores the full inner-product geometry of the variable vectors: diagonal entries are squared vector lengths and off-diagonal entries are the inner products from which angles and projections are derived. The retained matrix $M_J$ stores the geometry reproduced in the displayed subspace, whereas $H_J=M-M_J$ stores the omitted geometry. The bound therefore controls pairwise geometric distortion directly; high retained inertia alone does not guarantee that the particular relationships invoked by a claim are preserved.

\begin{remark}[Tolerance is substantive, not universal]
The analyst must justify $\mathcal P$, $d$, and $\tau$ from the scientific claim before using the display to support that claim. Sensitivity over plausible tolerances should be reported when no single cutoff is scientifically compelled.
\end{remark}

The methodological contribution is therefore not a new PCA algorithm. It is a target-explicit diagnostic protocol that changes how PCA-biplot analyses are specified, interpreted, and reported: the analyst must identify the represented operator, determine the identifiable spectral object, assess whether the retained projection supports the proposed scientific reading, and state whether the claim is well posed or ill posed and why.
 \subsection{Axes, invariant subspaces, and isotropy}
\label{subsec:spectral-objects}

Let
\[
M=Q\Lambda Q^{\top},\qquad \Lambda=\operatorname{diag}(\lambda_1,\ldots,\lambda_p),
\quad \lambda_1\geq\cdots\geq\lambda_p\geq0.
\]
If $\lambda_j$ is simple, its eigenvector is identifiable up to sign. This sign ambiguity is harmless for biplot geometry because simultaneous reversal of an axis and all coordinates on that axis preserves inner products, distances, and angles. By contrast, if an eigenvalue has multiplicity greater than one, the corresponding eigenspace is identifiable but its internal orthonormal basis is not. If $Q_J$ is one basis for a repeated-eigenvalue block and $H$ is any orthogonal matrix of the same dimension, then $Q_JH$ is an equally valid basis. Axis-specific loadings and projected angles that depend on a particular internal basis are therefore not population-defined quantities. The invariant object is the spectral projector
\[
P_J=Q_JQ_J^{\top},
\]
which is unchanged under $Q_J\mapsto Q_JH$.

For example, if $\lambda_1>\lambda_2=\lambda_3>\lambda_4$, then PC1 and PC4 are individually identifiable, whereas PC2 and PC3 are not. The plane $\operatorname{span}(q_2,q_3)$ is an identifiable invariant subspace; its coordinate axes are arbitrary. Classical perturbation theory further shows that small spectral separation makes empirical axes or subspaces sensitive to sampling or numerical perturbation \citep{DavisKahan1970,Wedin1972,StewartSun1990,YuWangSamworth2015}. Under full isotropy, $M=cI_p$, no individual axis or proper leading principal subspace is selected by the target. Expanded spectral details are provided in \SMfour{} (\SMfourTitle{}) \citep{Jolliffe1989}.
 \subsection{Projection adequacy and angular interpretation}
\label{subsec:projection}

Under variable principal coordinates, the full-dimensional variable vectors satisfy $GG^{\top}=M$, so $g_j^{\top}g_k=M_{jk}$ and
\[
\cos(\theta_{jk})=\frac{M_{jk}}{\sqrt{M_{jj}M_{kk}}}
\]
whenever $M_{jj}M_{kk}>0$. In covariance-PCA, vector lengths and inner products retain variance and covariance information, while the angle normalises by marginal vector lengths. In correlation-PCA, the same cosine is already a relationship on the standardised scale. These two readings are not interchangeable: they answer different target questions even when the biplot convention is the same.

If only a component set $J$ is displayed, define the projected Gram matrix
\[
K_J=G_JG_J^{\top}.
\]
For the usual PC1--PC2 display, $J=\{1,2\}$ and $K_J$ is the matrix of visible inner products among the projected variable vectors. For variables $j$ and $k$,
\[
g_j^{\top}g_k=g_{j,J}^{\top}g_{k,J}+g_{j,J^c}^{\top}g_{k,J^c},
\]
so the displayed biplot represents only the first term. The complementary term is the omitted inner product; it can be positive, negative, or zero, and may either reinforce or cancel the visible relationship. Consequently, a projected angle can be visually compelling while being a poor approximation to the full target relationship. If either projected vector has zero norm, the projected angle is undefined and must not be interpreted as evidence of orthogonality or absence of association.

Projection adequacy is therefore evaluated for the relationships invoked by the claim, not by retained inertia alone. For a displayed component set $J$, define
\[
e_{jk}(J)=\left|M_{jk}-g_{j,J}^{\top}g_{k,J}\right|,
\qquad
 e^{\mathrm{rel}}_{jk}(J)=\frac{e_{jk}(J)}{\sqrt{M_{jj}M_{kk}}}
\]
when the denominator is non-zero. For a prespecified pair set $\mathcal P$, the conservative claim-level diagnostic is
\[
E_{\max}(J;\mathcal P)=\max_{(j,k)\in\mathcal P}e^{\mathrm{rel}}_{jk}(J).
\]
When the claim is angular, the corresponding full-space and projected cosines should also be reported, because a small absolute inner-product error may still alter the angle substantially when projected vector lengths are small. The pair set, discrepancy measure, error values, and tolerance must be declared before the display is used as evidence. Further details on projected Gram matrices, angular diagnostics, and omitted-coordinate bounds are given in \SMthree{} and \SMfour{}.

\subsection{Operational diagnostic procedure}
\label{subsec:algorithm}

The formal definitions lead to the following reproducible diagnostic procedure, ending in a claim-specific classification as representationally well posed or ill posed.
\begin{enumerate}[label=(\arabic*)]
\item State the scientific claim, declare its target, and identify the operator and biplot convention used to represent that target.
\item Document and justify target alignment by comparing the declared phenomenon and its invariance requirements with the matrix actually diagonalised.
\item Inspect eigenvalue multiplicities and relevant eigengaps, and determine whether the claim may refer to an identifiable axis, an identifiable invariant subspace or projector, or no axis-specific population object.
\item For the prespecified claim-relevant relationships, compare the full-space and displayed quantities with the stated tolerance, then retain, reformulate, qualify, redirect, or reject the claim according to the failed condition or conditions.
\end{enumerate}

The procedure is reproducible even though the scientific target and tolerance require substantive judgement. The R and Python implementations compute the spectral and projection diagnostics required by steps (3)--(4); the expanded checklist, implementation details, and additional cases are documented in \SMfour{} (\SMfourTitle{}).

\Cref{tab:failure_modes} summarises the operational meaning of the three failure modes.

\begin{table}[htbp]
\centering
\caption{Operational interpretation of the three claim-admissibility failure modes.}
\label{tab:failure_modes}
\small
\begin{tabularx}{\textwidth}{>{\raggedright\arraybackslash}p{2.8cm} >{\raggedright\arraybackslash}p{3.4cm} >{\raggedright\arraybackslash}p{2.7cm} X}
\toprule
Failure mode & Diagnostic question & Consequence & Recommended action \\
\midrule
Target misalignment & Does the diagonalised matrix represent the scientific target? & The display answers a different scale-dependent question. & Redefine the target or justify the preprocessing before interpretation. \\
Spectral non-identifiability & Is the invoked axis unique, or is only an invariant subspace identified? & Axis-specific loadings and narratives may depend on an arbitrary basis. & Interpret a projector or invariant subspace, or avoid axis-specific claims. \\
Projection inadequacy & Does the displayed component set preserve the relationships required by the claim within tolerance $\tau$? & Displayed angles or products distort the relevant full-space geometry. & Change the displayed subspace, report the error, or qualify/reject the claim. \\
\bottomrule
\end{tabularx}
\end{table}

The decision is not merely classificatory: it determines what may responsibly be written from the display. \Cref{tab:claim_actions} summarises the claim-level actions that follow from the diagnosis.

\begin{table}[htbp]
\centering
\caption{Claim-level actions following the admissibility assessment. The action is conditional on the declared target, relation set, discrepancy, and tolerance.}
\label{tab:claim_actions}
\small
\begin{tabularx}{\textwidth}{>{\raggedright\arraybackslash}p{3.0cm} >{\raggedright\arraybackslash}p{4.1cm} X}
\toprule
Diagnostic outcome & Status of proposed statement & Permissible response \\
\midrule
All three conditions satisfied & Representationally well posed for the declared specification. & Retain the claim and report the target, spectral object, displayed component set, diagnostics, and tolerance. \\
Correct target; internal axes non-identifiable & Named-axis claim ill posed, but a complete invariant-subspace claim may be well posed. & Replace PC-axis narratives by projector- or subspace-level statements. \\
Correct target and identifiable object; inadequate projection & The selected display does not reproduce the relationships required by the claim. & Use another component set, increase dimensionality, report the discrepancy, or qualify the statement as display-specific. \\
Target misaligned & The biplot may be valid for another target, but the proposed scientific claim is ill posed for this operator. & Reanalyse the operator required by the question or rewrite the claim on the represented scale. \\
\bottomrule
\end{tabularx}
\end{table}

The expanded checklist is supplied in \SMfour{}.

 \subsection{Population-level scope and the methodological role of controlled scenarios}
\label{subsec:controlled-necessity}

The framework is population-based: it audits the representational properties of a declared target matrix, its identifiable spectral objects, and the loss induced by projection. Controlled scenarios are used because the target, spectrum, full-dimensional relationships, and projection error are exactly known. They are not intended to mimic sampling variability; they isolate the representational truth that a sample analysis is attempting to estimate.

A real-data example can illustrate application, but cannot by itself validate population representational truth because the underlying target geometry is unobserved. Sampling uncertainty, near-multiplicity, finite-sample calibration, high-dimensional regimes, and broader empirical validation are therefore subsequent inferential questions. In practice, finite-sample diagnostics should be reported conditionally on the declared target and claim: the Iris application in \SMfive{} shows how bootstrap stability information may accompany, but not replace, target specification and population diagnosis.

Correlation-PCA and standardised biplots remain appropriate when the scientific objective is explicitly correlation-scale, the relevant spectral objects are identifiable or treated as subspaces, and the displayed dimensions adequately represent the relationships of interest. The overextension examined here is narrower: reading a standardised low-dimensional biplot as evidence about raw-scale covariance or about relationships that the selected plane does not preserve.
  \section{Controlled Population Contrasts}
\label{sec:controlled-contrasts}

\subsection{Construction and coordinate convention}
\label{subsec:construction}

Let $X=(X_1,X_2,X_3,X_4)^{\top}$ have means $(10,50,200,500)$ and, unless otherwise stated, marginal variances $(5,20,50,80)$. The scenarios are population constructions, so the matrices analysed below are exact operators rather than estimates. Association refers throughout to linear association.

For eigenvalues $\lambda_a,\lambda_b$ and eigenvector entries $q_{ja},q_{jb}$, the displayed variable vector in plane $(\mathrm{PC}a,\mathrm{PC}b)$ is
\[
v_j^{(a,b)}=\left(\sqrt{\lambda_a}q_{ja},\sqrt{\lambda_b}q_{jb}\right).
\]
This formula fixes the coordinate convention used in the figures: changes in the target matrix or in the identifiable spectral object change the displayed vectors and therefore the admissible interpretation.

All coordinates and angle annotations for the main displays are reproduced by the parallel R and Python workflows supplied with the \MainRepo{}. \SMtwo{} provides the complete six-plane atlas and all pairwise diagnostics, together with its own supplement-specific reproducibility materials.
 \subsection{Null association: Scenarios A and B}
\label{subsec:AB}

\begin{figure}[!htbp]
\centering
\input{figures/Figure_AB_Null_Association.tex}
\caption{Scenario A versus Scenario B. Left: covariance-PCA under unequal independent variances, with identifiable coordinate axes. Right: correlation-PCA after unit-variance standardisation, where the isotropic target makes the displayed orientation arbitrary. Complete coordinates, diagnostics, and R/Python scripts are supplied with \SMtwo{}.}
\par\noindent\textbf{Alt text:} Two side-by-side variable-vector biplots compare covariance-PCA under unequal independent variances with correlation-PCA after standardisation. The covariance panel has ordered coordinate directions; the correlation panel is isotropic and shows one arbitrary orthonormal orientation.
\label{fig:A_B_PC1PC2}
\end{figure}
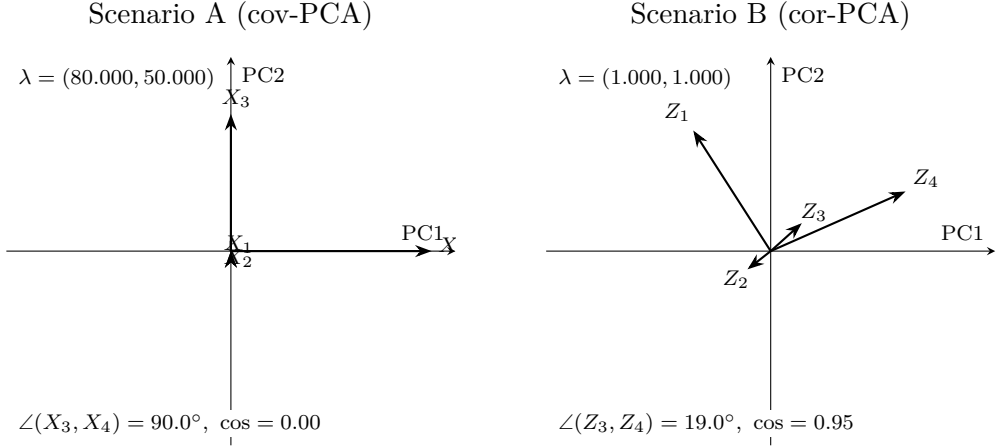

Scenario A uses
\[
\Sigma_A=\operatorname{diag}(5,20,50,80).
\]
Its eigenvalues are distinct, so all population axes are identifiable. As shown in the left panel of \cref{fig:A_B_PC1PC2}, the PC1--PC2 plane displays the dominant raw-scale variance directions $X_4$ and $X_3$ as an orthogonal pair. For claims restricted to its covariance target and displayed identifiable axes, the display is representationally well posed, although it is necessarily a partial representation of the four-dimensional structure.

Scenario B applies unit-variance standardisation to the same null-association construction, giving
\[
R_B=I_4.
\]
The correlation target is fully isotropic. Every orthonormal basis is admissible, so no individual PC axis or axis-specific two-dimensional geometry is population-identifiable. The right panel of \cref{fig:A_B_PC1PC2} shows one admissible displayed configuration, but alternative orthonormal bases can produce different acute or obtuse projected angles even though all off-diagonal target correlations are zero.

The contrast in \cref{fig:A_B_PC1PC2} isolates the central distinction: standardisation does not create association, but it can replace a distinct-spectrum covariance target by an isotropic correlation target whose axis-specific biplot geometry is arbitrary. The complementary planes and alternative admissible orientations are documented in \SMtwo{} and \SMfour{}.
 \subsection{Control for the null-association contrast: Scenario C}
\label{subsec:C-control}

Scenario C uses the raw-covariance target
\[
\Sigma_C=cI_4,\qquad c=\sqrt{50\times80}.
\]
It is therefore a covariance-scale control, not a correlation-PCA construction. Its complete spectrum is repeated, so the target identifies the full isotropic space but no individual principal axis or proper leading principal subspace. Under two equally admissible orthonormal bases, the same matrix can produce different projected vectors and angles; the change is generated solely by the arbitrary internal basis and is documented in \SMtwo{}.

Scenario C has a different methodological role from Scenario B. In Scenario B, routine standardisation transforms an unequal-variance covariance target into an isotropic correlation target. In Scenario C, isotropy is already present in the declared covariance target itself. Both cases make named-axis claims ill posed, but for different reasons in the audit: Scenario B illustrates target change under preprocessing, whereas Scenario C is a positive degeneracy control showing that axis non-identifiability is a property of the target spectrum, not a defect of the biplot algorithm or of correlation-PCA specifically. Scenario C therefore separates absence of association from axis identifiability: a diagonal matrix with unequal entries, as in Scenario A, identifies ordered coordinate axes, whereas a scalar multiple of the identity does not.

The corresponding two-basis display, complete coordinates, pairwise angles, and alternative-basis diagnostics are provided in \SMtwo{}, with expanded projector-level interpretation in \SMfour{}.
 \subsection{Positive association: Scenarios D and E}
\label{subsec:DE}

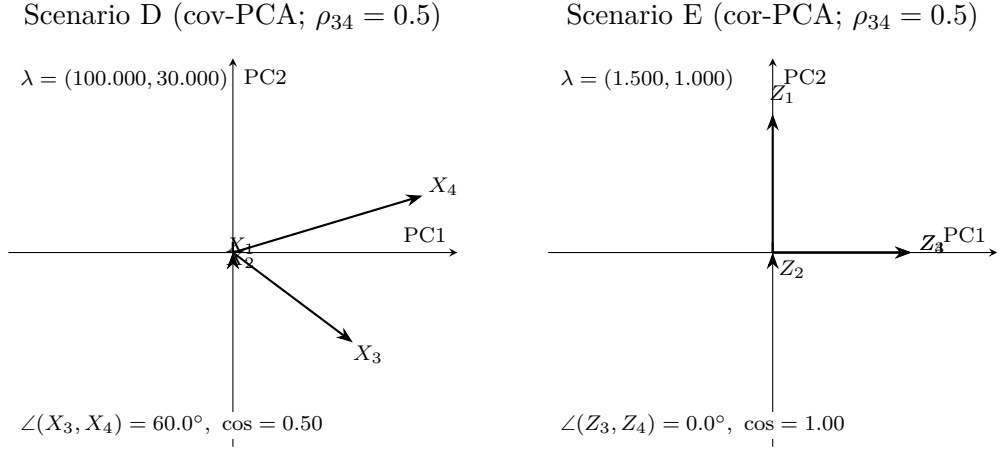
\begin{figure}[!htbp]
\centering
\input{figures/Figure_DE_Positive_Association.tex}
\caption{Scenario D versus Scenario E. Left: covariance-PCA under positive linear association in $(X_3,X_4)$. Right: correlation-PCA after unit-variance standardisation; the PC1--PC2 display collapses the focal projected angle to $0^\circ$ because the omitted contrast direction is required for the full $60^\circ$ relationship. Complete coordinates, diagnostics, and R/Python scripts are supplied with \SMtwo{}.}
\par\noindent\textbf{Alt text:} Two side-by-side variable-vector biplots compare covariance- and correlation-scale representations of a positive association between variables 3 and 4. The right panel shows coincident projected vectors in PC1--PC2 although the full-space angle is 60 degrees.
\label{fig:D_E_PC1PC2}
\end{figure}

Scenario D retains the heterogeneous marginal variances and imposes $\Corr(X_3,X_4)=0.5$:
\[
\Sigma_{D}=
\begin{pmatrix}
5&0&0&0\\
0&20&0&0\\
0&0&50&\sqrt{50\times80}/2\\
0&0&\sqrt{50\times80}/2&80
\end{pmatrix}.
\]
Its eigenvalues are $(100,30,20,5)$, so the individual axes are identifiable. The leading plane, shown in the left panel of \cref{fig:D_E_PC1PC2}, represents the associated high-variance block on the raw covariance scale.

After standardisation, Scenario E has
\[
R_{E}=
\begin{pmatrix}
1&0&0&0\\
0&1&0&0\\
0&0&1&0.5\\
0&0&0.5&1
\end{pmatrix},
\]
with spectrum $(1.5,1,1,0.5)$. PC1 and PC4 are identifiable, whereas PC2 and PC3 are identifiable only through their two-dimensional invariant subspace. In the PC1--PC2 plane shown in the right panel of \cref{fig:D_E_PC1PC2}, $Z_3$ and $Z_4$ coincide because the identifiable contrast direction in PC4 is omitted. The projected angle is therefore $0^{\circ}$ although their full-space variable-principal-coordinate angle is $60^{\circ}$.

Scenario E is consequently representationally well posed for claims about its correlation target and identifiable common direction, but the PC1--PC2 claim that $Z_3$ and $Z_4$ have a $0^\circ$ full-space relationship is ill posed because \cref{fig:D_E_PC1PC2} is not a complete angular map of $R_{E}$. The PC1--PC4 plane, shown in \SMtwo{}, recovers the full $60^{\circ}$ focal relationship because it includes both the common and contrast directions.
 \subsection{Control for the positive-association contrast: Scenario F}
\label{subsec:F-control}

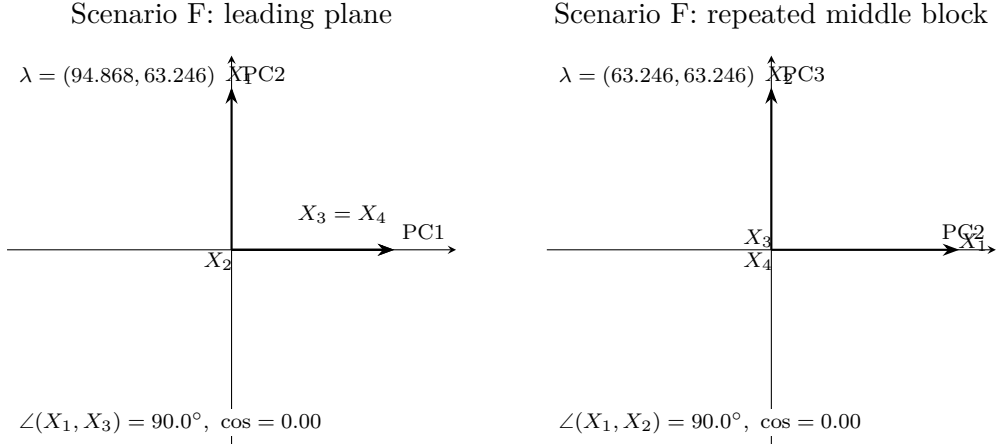
\begin{figure}[!htbp]
\centering
\input{figures/Figure_F_Partial_Identifiability.tex}
\caption{Scenario F, a raw-covariance target with equal marginal variances and positive association between $X_3$ and $X_4$. Left: PC1--PC2 combines the identifiable leading direction with one basis-dependent axis from the repeated middle block. Right: PC2--PC3 displays the repeated-eigenvalue plane under one admissible basis. Scenario F supports well-posed claims for its covariance target, identifiable leading axis, identifiable trailing axis, and invariant middle subspace; separate PC2 and PC3 narratives are ill posed within the repeated block. Complete diagnostics are supplied with \SMtwo{} and \SMfour{}.}
\par\noindent\textbf{Alt text:} Two biplots for a covariance matrix with equal marginal variances and positive association. The left PC1--PC2 panel combines an identifiable leading direction with one basis-dependent middle direction; the right PC2--PC3 panel displays the repeated invariant subspace under one admissible basis.
\label{fig:F_leading_repeated}
\end{figure}

Scenario F has equal marginal variances and the same $X_3$--$X_4$ association as Scenario D; its spectrum is $(1.5c,c,c,0.5c)$. It identifies the leading sum direction, the trailing contrast direction, and the middle invariant subspace, but not the individual PC2 and PC3 axes. Hence multiplicity does not invalidate the entire decomposition: it changes the object of interpretation from selected axes to the invariant subspace. \Cref{fig:F_leading_repeated} illustrates two distinct levels of interpretation. Its left panel is the routine PC1--PC2 display, which combines an identifiable leading direction with one basis-dependent direction from the repeated block. Its right panel is PC2--PC3, the repeated invariant plane itself, shown under one admissible basis. Complete derivations, additional admissible bases, and related spectral controls are provided in \SMtwo{} and \SMfour{}.

 \subsection{Cross-scenario synthesis}
\label{subsec:cross-synthesis}

Taken together, Scenarios A--F separate four questions that should not be collapsed: which scientific target is intended, which matrix is diagonalised, which spectral object is identifiable, and whether the selected plane reproduces the relationships used in the interpretation. Representational well-posedness is therefore claim specific rather than a property of a PCA solution as a whole. Scenarios A and D support well-posed covariance claims when the declared target and displayed projection match the relationship invoked. Scenarios B and E are ill posed for raw-scale covariance claims after standardisation, although they may support claims explicitly formulated for their correlation targets. Scenario C is ill posed for named-axis interpretation because complete isotropy leaves the population eigendirections unidentified. Scenario F is well posed at the invariant-subspace level but ill posed for individual-axis interpretations within the repeated block. \Cref{tab:scenario-status} summarises these distinctions.

\begin{table}[htbp]
\centering
\caption{Cross-scenario representational synthesis. Status is conditional on the stated claim and displayed object.}
\label{tab:scenario-status}
\scriptsize
\begin{tabularx}{\textwidth}{>{\raggedright\arraybackslash}p{1.0cm} >{\raggedright\arraybackslash}p{4.1cm} X}
\toprule
Scenario & Target and identifiable object & Representational status \\
\midrule
A & Raw covariance; distinct individual axes & Well posed for declared covariance claims supported by the displayed plane. \\
B & Correlation after standardisation; no individual axis under full isotropy & Ill posed for axis-specific claims and for raw-scale covariance claims. \\
C & Isotropic raw covariance; full space only & Ill posed for named-axis and proper-leading-subspace interpretations. \\
D & Raw covariance with positive association; distinct individual axes & Well posed for declared covariance claims supported by the displayed plane. \\
E & Correlation after standardisation; leading axis and repeated-eigenvalue subspace & Well posed for appropriately stated correlation-scale claims; ill posed for raw-scale or incomplete-plane claims. \\
F & Raw covariance with a repeated middle eigenvalue; leading and trailing axes plus middle invariant subspace & Well posed at the identifiable-axis or subspace level; ill posed for separate PC2/PC3 narratives. \\
\bottomrule
\end{tabularx}
\end{table}
\section{Discussion}
\label{sec:discussion}

The motivating difficulty is that unit-variance standardisation is often a routine preprocessing step, although it replaces covariance geometry by correlation geometry. This replacement is legitimate when the scientific target is standardised association, but it changes the admissible content of a biplot interpretation. A statement about raw-scale morphology, absolute variation, or covariance structure cannot be carried over silently to a display whose represented operator is $R$.

The framework developed here makes this distinction operational. Target alignment asks whether the diagonalised matrix represents the phenomenon named in the claim. Spectral identifiability asks whether the language refers to a population-defined axis or only to an invariant subspace. Projection adequacy asks whether the displayed component set preserves the relationships on which the claim depends. These diagnoses convert an informal visual interpretation into an auditable statement with a specified remedy: retain the claim, reformulate it at subspace level, qualify it as plane-specific, redirect it to another display, or reject it for the declared target.

The controlled scenarios show why all three diagnoses are needed. Standardisation can turn a distinct-spectrum covariance target into an isotropic correlation target without creating association; incomplete projection can collapse a full-space $60^\circ$ relationship to $0^\circ$; and repeated eigenvalues can leave an invariant subspace identifiable while making named-axis claims arbitrary. The contrast among Scenarios C, E, and F is especially important. Scenario C is a degeneracy control: no named axis is population selected. Scenario E shows that even a legitimate correlation target can be misread when an incomplete plane is treated as a full-space angular map. Scenario F shows the intermediate case in which some spectral objects are identifiable but the individual axes inside a repeated block are not. The framework is therefore neither a criticism of correlation-PCA nor a general warning against biplots; it is a rule for matching the sentence to the target, the spectrum, and the displayed geometry.

This point is reinforced by the supplementary materials. The positive correlation-PCA control in \SMthree{} shows that correlation-PCA is representationally well posed when $R$ is the declared target and the claim is formulated at the identifiable subspace level. The Iris application in \SMfive{} shows how the same logic applies to real data: covariance-PCA supports claims about raw floral-organ morphology and absolute variation, whereas correlation-PCA supports claims about standardised association among traits. The two displays are complementary, not interchangeable.

The proposal is diagnostic rather than prohibitive. Standard PCA software may compute the eigendecomposition and biplot correctly, but implementation-level correctness does not establish scientific admissibility. The durable reporting sequence is therefore question first, target second, spectrum third, and display fourth.

The framework is also not a substitute for sampling inference, model checking, or domain-specific validation. It identifies whether a proposed interpretation is representationally licensed at the population-target level; empirical uncertainty must then be assessed conditionally on that target and claim. The population analysis supplied here provides the target object that later inferential procedures must estimate, while sampling stability, near-multiplicity, finite-sample calibration, and broader empirical validation remain design-specific tasks.

\section{Conclusions}
\label{sec:conclusion}

Unit-variance standardisation is not a neutral preliminary step in PCA-biplot analysis. It replaces covariance geometry by correlation geometry and therefore changes the operator whose spectrum, variable coordinates, and angular relationships are displayed. This change is valid when the scientific question concerns standardised association; it does not automatically support conclusions about raw-scale covariance, absolute variation, or relationships that the selected plane fails to preserve.

The central result is methodological: computational correctness is necessary but not sufficient for scientific interpretability. A PCA-biplot claim is representationally well posed only when the represented operator matches the declared scientific target, the invoked axis or invariant subspace is identifiable at the level claimed, and the displayed projection preserves the relationships on which the conclusion depends. Failure of any condition does not make PCA invalid. It identifies the specific way in which the proposed scientific sentence must be reformulated, qualified, redirected, or withdrawn.

The framework therefore improves PCA-biplot reporting by defining the evidence required to move from a correct decomposition to a scientific sentence. Covariance-PCA and correlation-PCA are complementary when their targets are declared: the former is suited to claims about raw-scale covariance and absolute variation, whereas the latter is suited to claims about standardised association. They become problematic only when a claim formulated for one target is silently transferred to the other, or when a low-dimensional display is asked to support relationships that it does not preserve. A minimal report should therefore state the target, the spectral object, the displayed component set, the claim-relevant relationships, and the projection diagnostic used to license the interpretation.

\section*{Data and Code Availability}

All controlled population matrices are defined in the manuscript and supplementary materials. The reproducibility materials are organised as a main-manuscript package and supplement-specific packages. The \MainRepo{} contain the LaTeX source, figure files, R/Python scripts, CSV outputs, software requirements, and verification notes for the results reported in the main text; each supplementary-material package contains the corresponding source files, scripts, data, generated outputs, and README documentation. Fisher's iris data \citep{Fisher1936} are included with the empirical-application materials as \texttt{iris\_data.csv}. The Python workflow for the main text was executed in a clean verification environment; the parallel R workflow is supplied for independent execution under R 4.3.3 or later \citep{RCoreTeam2024}.

The complete reproducibility package for the manuscript and supplementary materials is archived on Zenodo at \url{https://doi.org/10.5281/zenodo.21361296} (DOI: \href{https://doi.org/10.5281/zenodo.21361296}{10.5281/zenodo.21361296}). The archive contains the main-manuscript package, supplement-specific packages, LaTeX sources, figure files, R/Python scripts, CSV outputs, software requirements, README files, and verification notes required to reproduce the reported population scenarios, biplot diagnostics, projection diagnostics, and empirical application.

\section*{Author Information}

Luiz Roberto Martins Pinto is a retired professor of the Department of Exact and Technological Sciences, State University of Santa Cruz (UESC), Ilh\'eus, Bahia, Brazil. Carlos Tadeu dos Santos Dias is a retired professor of the Department of Exact Sciences, Luiz de Queiroz College of Agriculture (ESALQ), University of S\~ao Paulo (USP), Piracicaba, S\~ao Paulo, Brazil. Luiz Roberto Martins Pinto is the corresponding author.
 \section*{Disclosure Statement}

The authors report no conflict of interest.
 \section*{Funding}

No external funding was received for this work.
 \section*{Use of Generative Artificial Intelligence}

Generative artificial intelligence tools were used for language editing, structural review, consistency checking, and assistance with reproducibility documentation. The authors independently verified the mathematical arguments, numerical results, references, code, and final wording and take full responsibility for the content of the manuscript.

  \end{document}

%% file: figures/Figure_AB_Null_Association.tex
\begin{tikzpicture}
\begin{groupplot}[group style={group size=2 by 1, horizontal sep=1.2cm}]
\nextgroupplot[
  width=0.47\linewidth,height=0.42\linewidth,
  axis lines=middle, xtick=\empty, ytick=\empty,
  ticklabel style={font=\scriptsize}, label style={font=\scriptsize}, title style={font=\small},
  title={Scenario A (cov-PCA)}, xlabel={PC1}, ylabel={PC2},
  xmin=-10.0, xmax=10.0, ymin=-10.0, ymax=10.0,
]
\addplot[thick, -{Stealth[length=2.2mm,width=1.6mm]}] coordinates {(0,0) (0.00000,0.00000)};
\node[font=\scriptsize] at (axis cs:0.35000,0.35000) {$ X_1 $};
\addplot[thick, -{Stealth[length=2.2mm,width=1.6mm]}] coordinates {(0,0) (0.00000,0.00000)};
\node[font=\scriptsize] at (axis cs:0.35000,-0.35000) {$ X_2 $};
\addplot[thick, -{Stealth[length=2.2mm,width=1.6mm]}] coordinates {(0,0) (0.00000,7.07107)};
\node[font=\scriptsize] at (axis cs:0.25000,7.88675) {$ X_3 $};
\addplot[thick, -{Stealth[length=2.2mm,width=1.6mm]}] coordinates {(0,0) (8.94427,0.00000)};
\node[font=\scriptsize] at (axis cs:9.90981,0.25000) {$ X_4 $};
\node[font=\scriptsize,anchor=north west,fill=white,inner sep=1pt] at (rel axis cs:0.02,0.98) {$\lambda=(80.000,50.000)$};
\node[font=\scriptsize,anchor=south west,fill=white,inner sep=1pt] at (rel axis cs:0.02,0.02) {$\angle(X_3,X_4)=90.0^\circ,\ \cos=0.00$};

\nextgroupplot[
  width=0.47\linewidth,height=0.42\linewidth,
  axis lines=middle, xtick=\empty, ytick=\empty,
  ticklabel style={font=\scriptsize}, label style={font=\scriptsize}, title style={font=\small},
  title={Scenario B (cor-PCA)}, xlabel={PC1}, ylabel={PC2},
  xmin=-1.40, xmax=1.40, ymin=-1.40, ymax=1.40,
]
\addplot[thick, -{Stealth[length=2.2mm,width=1.6mm]}] coordinates {(0,0) (-0.48298,0.87001)};
\node[font=\scriptsize] at (axis cs:-0.58162,0.99962) {$ Z_1 $};
\addplot[thick, -{Stealth[length=2.2mm,width=1.6mm]}] coordinates {(0,0) (-0.14494,-0.13170)};
\node[font=\scriptsize] at (axis cs:-0.21653,-0.20223) {$ Z_2 $};
\addplot[thick, -{Stealth[length=2.2mm,width=1.6mm]}] coordinates {(0,0) (0.19435,0.20185)};
\node[font=\scriptsize] at (axis cs:0.26990,0.27800) {$ Z_3 $};
\addplot[thick, -{Stealth[length=2.2mm,width=1.6mm]}] coordinates {(0,0) (0.84140,0.43010)};
\node[font=\scriptsize] at (axis cs:0.96871,0.52451) {$ Z_4 $};
\node[font=\scriptsize,anchor=north west,fill=white,inner sep=1pt] at (rel axis cs:0.02,0.98) {$\lambda=(1.000,1.000)$};
\node[font=\scriptsize,anchor=south west,fill=white,inner sep=1pt] at (rel axis cs:0.02,0.02) {$\angle(Z_3,Z_4)=19.0^\circ,\ \cos=0.95$};
\end{groupplot}
\end{tikzpicture}

%% file: figures/Figure_DE_Positive_Association.tex
\begin{tikzpicture}
\begin{groupplot}[group style={group size=2 by 1, horizontal sep=1.2cm}]
\nextgroupplot[
  width=0.47\linewidth,height=0.42\linewidth,
  axis lines=middle, xtick=\empty, ytick=\empty,
  ticklabel style={font=\scriptsize}, label style={font=\scriptsize}, title style={font=\small},
  title={Scenario D (cov-PCA; $\rho_{34}=0.5$)}, xlabel={PC1}, ylabel={PC2},
  xmin=-10.0, xmax=10.0, ymin=-10.0, ymax=10.0,
]
\addplot[thick, -{Stealth[length=2.2mm,width=1.6mm]}] coordinates {(0,0) (0.00000,0.00000)};
\node[font=\scriptsize] at (axis cs:0.35000,0.35000) {$ X_1 $};
\addplot[thick, -{Stealth[length=2.2mm,width=1.6mm]}] coordinates {(0,0) (0.00000,0.00000)};
\node[font=\scriptsize] at (axis cs:0.35000,-0.35000) {$ X_2 $};
\addplot[thick, -{Stealth[length=2.2mm,width=1.6mm]}] coordinates {(0,0) (5.34522,-4.62910)};
\node[font=\scriptsize] at (axis cs:6.02284,-5.24943) {$ X_3 $};
\addplot[thick, -{Stealth[length=2.2mm,width=1.6mm]}] coordinates {(0,0) (8.45154,2.92770)};
\node[font=\scriptsize] at (axis cs:9.37767,3.41192) {$ X_4 $};
\node[font=\scriptsize,anchor=north west,fill=white,inner sep=1pt] at (rel axis cs:0.02,0.98) {$\lambda=(100.000,30.000)$};
\node[font=\scriptsize,anchor=south west,fill=white,inner sep=1pt] at (rel axis cs:0.02,0.02) {$\angle(X_3,X_4)=60.0^\circ,\ \cos=0.50$};

\nextgroupplot[
  width=0.47\linewidth,height=0.42\linewidth,
  axis lines=middle, xtick=\empty, ytick=\empty,
  ticklabel style={font=\scriptsize}, label style={font=\scriptsize}, title style={font=\small},
  title={Scenario E (cor-PCA; $\rho_{34}=0.5$)}, xlabel={PC1}, ylabel={PC2},
  xmin=-1.40, xmax=1.40, ymin=-1.40, ymax=1.40,
]
\addplot[thick, -{Stealth[length=2.2mm,width=1.6mm]}] coordinates {(0,0) (0.00000,1.00000)};
\node[font=\scriptsize] at (axis cs:0.06000,1.14000) {$ Z_1 $};
\addplot[thick, -{Stealth[length=2.2mm,width=1.6mm]}] coordinates {(0,0) (0.00000,0.00000)};
\node[font=\scriptsize] at (axis cs:0.12000,-0.12000) {$ Z_2 $};
\addplot[thick, -{Stealth[length=2.2mm,width=1.6mm]}] coordinates {(0,0) (0.86603,0.00000)};
\node[font=\scriptsize] at (axis cs:0.99531,0.06000) {$ Z_3 $};
\addplot[thick, -{Stealth[length=2.2mm,width=1.6mm]}] coordinates {(0,0) (0.86603,0.00000)};
\node[font=\scriptsize] at (axis cs:0.99531,0.06000) {$ Z_4 $};
\node[font=\scriptsize,anchor=north west,fill=white,inner sep=1pt] at (rel axis cs:0.02,0.98) {$\lambda=(1.500,1.000)$};
\node[font=\scriptsize,anchor=south west,fill=white,inner sep=1pt] at (rel axis cs:0.02,0.02) {$\angle(Z_3,Z_4)=0.0^\circ,\ \cos=1.00$};
\end{groupplot}
\end{tikzpicture}

%% file: figures/Figure_F_Partial_Identifiability.tex
\begin{tikzpicture}
\begin{groupplot}[group style={group size=2 by 1, horizontal sep=1.2cm}]
\nextgroupplot[
  width=0.47\linewidth,height=0.42\linewidth,
  axis lines=middle, xtick=\empty, ytick=\empty,
  ticklabel style={font=\scriptsize}, label style={font=\scriptsize}, title style={font=\small},
  title={Scenario F: leading plane}, xlabel={PC1}, ylabel={PC2},
  xmin=-9.5,xmax=9.5,ymin=-9.5,ymax=9.5,
]
\addplot[thick,-{Stealth[length=2.2mm,width=1.6mm]}] coordinates {(0,0) (0,7.95271)};
\node[font=\scriptsize] at (axis cs:0.35,8.55) {$X_1$};
\node[font=\scriptsize] at (axis cs:-0.55,-0.55) {$X_2$};
\addplot[thick,-{Stealth[length=2.2mm,width=1.6mm]}] coordinates {(0,0) (6.88725,0)};
\node[font=\scriptsize,anchor=south,fill=white,inner sep=1pt] at (axis cs:4.7,1.15) {$X_3=X_4$};
\addplot[thick,-{Stealth[length=2.2mm,width=1.6mm]}] coordinates {(0,0) (6.88725,0)};
\node[font=\scriptsize,anchor=north west,fill=white,inner sep=1pt] at (rel axis cs:0.02,0.98) {$\lambda=(94.868,63.246)$};
\node[font=\scriptsize,anchor=south west,fill=white,inner sep=1pt] at (rel axis cs:0.02,0.02) {$\angle(X_1,X_3)=90.0^\circ,\ \cos=0.00$};

\nextgroupplot[
  width=0.47\linewidth,height=0.42\linewidth,
  axis lines=middle, xtick=\empty, ytick=\empty,
  ticklabel style={font=\scriptsize}, label style={font=\scriptsize}, title style={font=\small},
  title={Scenario F: repeated middle block}, xlabel={PC2}, ylabel={PC3},
  xmin=-9.5,xmax=9.5,ymin=-9.5,ymax=9.5,
]
\addplot[thick,-{Stealth[length=2.2mm,width=1.6mm]}] coordinates {(0,0) (7.95271,0)};
\node[font=\scriptsize] at (axis cs:8.55,0.35) {$X_1$};
\addplot[thick,-{Stealth[length=2.2mm,width=1.6mm]}] coordinates {(0,0) (0,7.95271)};
\node[font=\scriptsize] at (axis cs:0.35,8.55) {$X_2$};
\node[font=\scriptsize] at (axis cs:-0.55,0.55) {$X_3$};
\node[font=\scriptsize] at (axis cs:-0.55,-0.55) {$X_4$};
\node[font=\scriptsize,anchor=north west,fill=white,inner sep=1pt] at (rel axis cs:0.02,0.98) {$\lambda=(63.246,63.246)$};
\node[font=\scriptsize,anchor=south west,fill=white,inner sep=1pt] at (rel axis cs:0.02,0.02) {$\angle(X_1,X_2)=90.0^\circ,\ \cos=0.00$};
\end{groupplot}
\end{tikzpicture}